\documentstyle[psfig]{aipproc}

\begin{document}

\title{THE EMERGENCE OF A BLACK HOLE IN SUPERNOVA EXPLOSIONS}
 
\author{L. Zampieri$^{*}$, M. Colpi$^{\dagger}$,
S. Shapiro$^{*,\dagger \dagger}$, and I. Wasserman$^{\dagger \dagger \dagger}$}
\address{$^*$Department of Physics, University of Illinois at
Urbana--Champaign, Urbana, IL 61801\\
$^\dagger$Dipartimento 
di Fisica, Universit\`a degli Studi di
Milano, I--20133 Milano, Italy\\
$^{\dagger \dagger}$Department of Astronomy and National Center for
Supercomputing Applications, University of Illinois at
Urbana--Champaign, Urbana, IL 61801\\
$^{\dagger \dagger \dagger}$Center for Radiophysics and Space Research, Cornell
University, Ithaca, NY 14853}

\maketitle

\begin{abstract}

We present results from a fully
relativistic investigation of matter fallback in a supernova, assuming
that the central compact star is a black hole (BH).
Our goal is to answer the following questions:
can we infer from the light curve whether a stellar black hole (BH)  has 
formed in the 
aftermath of a supernova explosion ?  Did SN1897A produce a BH and, if so, 
when will the hole become detectable ?
We show that it should not be possible to infer the presence of a BH in
the remnant for another several hundred to a thousand years.
In the case of SN1987A, we estimate the BH luminosity to be
$L\simeq 5\times 10^{34}$ erg s$^{-1}$, which is
well below the present day
bolometric luminosity of the remnant ($\sim 10^{36}{\rm erg\,s^{-1}}$;
Suntzeff 1997\cite{Sun97}).

\end{abstract}

\section*{Introduction}

Stellar evolution calculations show that stars with a
main sequence mass in the range 8--19 $M_\odot$ finish their lives
with a compact core of $\sim 1.4 M_\odot$
(Woosley \& Weaver 1995 \cite{WoosWea95}). For these stars the amount of 
material that falls back toward the core in the aftermath of its collapse
is negligible, so it is likely that they give birth to neutron stars.
The fate of stars with main sequence mass in the
range 19--25 $M_\odot$ is far less obvious. At the end of their
evolution, they have core masses around 1.6--1.8 $M_\odot$ and a   
variable amount of matter, 0.1--0.3 $M_\odot$ may fall back.
Typically, these explosions leave a central compact remnant of
1.7--2.1 $M_\odot$ whose nature depends critically on the maximum
(gravitational) mass
of a neutron star, $M_{crit}$, and hence on the equation of state of
dense nuclear matter.
To date, the value of $M_{crit}$ is quite uncertain:
$1.5 M_\odot < M_{crit} < 2.5 M_\odot$ for a nonrotating star.

The mass of the progenitor of SN1987A
($M =$ 18--21 $M_\odot$) falls in the range of masses where the outcome
of core collapse is uncertain;
from a theoretical
point of view, we do not know if a neutron star or a black hole
formed during the explosion.
However, if a compact object of any kind is present,
it is probably accreting from the progenitor stellar material
(Chevalier 1989\cite{Chev89}).
The bolometric light curve observed to date
can be explained by the ``standard''
theory of Type II supernovae.
It is therefore important to determine what extra luminosity
an accreting central component would produce and when its
presence might be discernible.
In this paper we present results from the first fully
relativistic investigation of supernova fallback in presence of a BH
including the effects of hydrodynamics and radiation transport.

\section*{Method and numerical results }

We constructed a spherically symmetric, general--relativistic
radiation hydrodynamic Lagrangian code capable of 
handling  the transfer
of radiation from the early  phase
when photons diffuse through a high temperature, expanding
cloud, to the late stage when the hydrogen envelope has
recombined and most of the ejecta are transparent
(Zampieri et al. (1997)\cite{Zam97}).
We consider fallback  from  a hydrogen ``cloud"; all of the emitted
energy comes from the release of heat residing in the gas originally,
or is generated by compressional heating in the course of accretion onto
the BH. Radioactive energy sources have not
been included in our calculations.
At the onset of evolution the cloud has homogenous density $\rho_0$
and a velocity profile which scales linearly with radius $r$.
The temperature profile is taken to be equal 
to the ``radiative zero solution'' given by Arnett (1980)\cite{Arn80} and the 
degree of ionization is computed throughout evolution according to the Saha 
equation. Four parameters uniquely specify the state of the ``cloud'':
the total mass $M_{cloud}$, the radius $r_{out}$,
the sound speed at the inner boundary $c_{s,0}$, and the ratio
${\tilde k}$ of the
accretion timescale $t_{a,0} = GM_{bh}/c_{a,0}^3$ ($M_{bh}$ mass of the BH)
to the expansion time $t_0 = r_{out}/V_0$ ($V_0$ maximum expansion 
velocity of the ejecta).

\begin{figure} 
\centerline{\psfig{figure=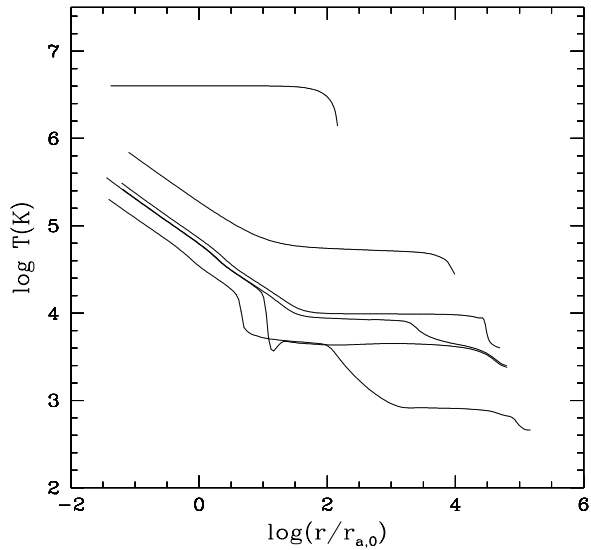,width=3.2in,height=3.0in}}
\vspace{10pt}
\caption{Gas temperature $T$ versus radius $r$ (in units of the initial
accretion radius $r_{a,0} = GM_{bh}/c_{s,0}^2$) at selected times for model A.
From the top to the bottom: $t/t_{a,0}$ = 0, 729, 3757, 4755, 4870, 11175.
}\label{myfirstfigure}
\end{figure}

We find two phases of evolution. In the first stage,
gas expands adiabatically in the outer region and accretes close to the
central hole. During this phase, the cloud is radiation dominated.
A second phase follows, during which hydrogen recombines and the ``cloud''
becomes optically thin. Fig.~\ref{myfirstfigure} shows 
the temperature profile of model A
characterized by $M_{cloud}=1 M_{\odot}$ ($\rho_0=2\times 
10^{-5}$ g cm$^{-3}$), $r_{out} = 3 \times 10^{12}$ cm,
${\tilde k} = 0.1$ and $c_{s,0}=10^8$ cm s$^{-1}$.
A recombination 
wave develops that propagates rapidly through the envelope.
The recombination front stalls as soon as it approaches the innermost region
where compressional heating due to accretion overcomes radiative cooling.
Outside the recombination front the flow is transparent but it remains
optically thick inside for a long time.
The density and velocity profiles display a self--similar character and are 
unaffected by the propagation of the recombination front. 


As recombination begins, the  huge amount of internal energy released gives 
rise to a big 
bump in the light curve at $t \sim 10$ days, as illustrated in
Fig.~\ref{mysecondfigure}.
After the emission maximum, the luminosity falls off abruptly. 
During the late evolution, the light curve is entirely powered by
accretion onto the central BH and the luminosity decreases
as a power law with time.
At late times, 
the evolution proceeds through a 
sequence of quasistationary states, and  $\dot {m}$ (in Eddington units) can
be estimated using the scaling relation for a pressureless fluid
(Colpi et al. 1996\cite{Col96}; equation [29])
\begin{equation}
\label{dustmdot}
{\dot {m}}(t)\simeq [4 \pi^{2/3}/9]
\rho_0 \,t_0 k_{es} c
\left ({t/t_0 }\right )^{-5/3} \, ,
\end{equation}
where $k_{es}$ is the electron scattering opacity.
The luminosity $l$ (in 
Eddington units) is close to that derived by Blondin (1986)\cite{Blon86} 
for stationary accretion
\begin{equation}
l \simeq 3 \times 10^{-7} 
\left({M_{bh}}/{M_\odot}\right)^{-1/3}
{\dot m}^{5/6}.  
\label{lumblon}
\end{equation}
In Fig.~\ref{mysecondfigure}, the dotted line at the bottom denotes the 
late--time accretion luminosity estimated using equations (\ref{dustmdot}) and
(\ref{lumblon}): the analytic extrapolation is in good agreement
with the computed light curve.

\section*{The emergence of a BH in SN1987A}

\begin{figure} 
\centerline{\psfig{figure=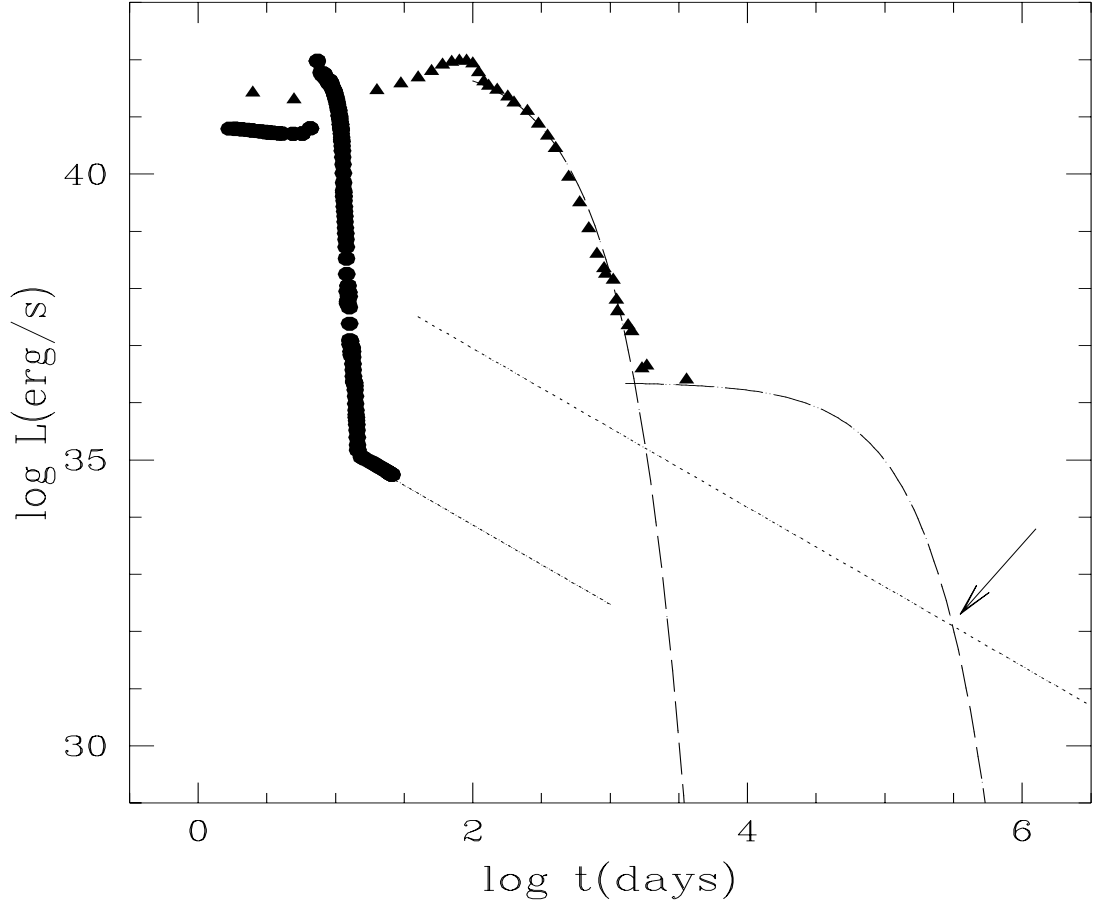,width=6.0in,height=5.0in}}
\vspace{10pt}
\caption{
The {\it Circles} give the solution for model A.
The {\it Dotted line} is the extrapolation
of the late-time evolution.
The {\it Triangles} are the bolometric luminosity of SN1987A.
The {\it Dashed lines} represent the contribution from the decay
of radioactive elements (0.07 $M_\odot$ of
$^{56}$Co and $\sim 5 \times 10^{-5} M_\odot$ of $^{44}$Ti).
Finally, the {\it upper dotted line} denotes the expected bolometric
luminosity emitted by a putative BH in SN1987A (equation [3]).
The {\it arrow} marks the time ($t \simeq 900$ years) of the
BH emergence.
}\label{mysecondfigure}
\end{figure}

Slightly after maximum, SN1987A
was powered by the radioactive decay of $^{56}$Co and  $^{57}$Co, and
at present its light curve is
consistent with emission from the decay of $^{44}$Ti
(Suntzeff 1997\cite{Sun97}; Fig.~\ref{mysecondfigure}).
Can we discern the luminosity emitted by the
accreting BH above the contribution resulting from $^{44}$Ti ?
An upper limit to the accretion luminosity of SN1987A can be inferred from
equations (\ref{dustmdot}) and (\ref{lumblon})
(for details see Zampieri et al. (1997)\cite{Zam97}). Following Chevalier
(1989)\cite{Chev89}, we adopt $\rho_0\, t_0^3\approx 10^9$ cgs
and $t_0\approx 7000$ seconds, yielding a luminosity
\begin{equation}
\label{lumfall}
l \simeq {8\times 10^{-3}\over (M_{bh}/M_\odot)^{1/3}[t({\rm years})]^{25/18}} 
\, .
\end{equation}
Equation (\ref{lumfall}) implies $L\simeq 5\times 10^{34}{\rm erg\,s^{-1}}$
after $t\approx 10$ years, which is well below the present day
bolometric luminosity of the remnant ($\sim 10^{36}{\rm erg\,s^{-1}}$;
Suntzeff 1997\cite{Sun97}) and also smaller than the luminosity estimated 
to result from radioactive decay. Thus, there is no observation that rules 
out the possibility that a BH resides inside the SN1987A remnant.
In Fig.~\ref{mysecondfigure} we plot the late--time light curve
of SN1987A calculated using (\ref{lumfall}).
As radioactive decay plummets at around 270--2700 years, it is clear that 
the BH would only appear after about 900 years irrespective of the 
detailed numbers.
After this time has elapsed, the luminosity of the remnant would be
$\sim 10^{32}{\rm erg\,s^{-1}}$, too dim for detection with present
technology. The inner part of the accretion flow
will be still optically thick and will be emitting roughly a black body
spectrum peaked in the visible band.


\end{document}